\documentclass[conference]{IEEEtran}
\IEEEoverridecommandlockouts
\usepackage{array}
\usepackage{makecell}
\usepackage{listings}
\usepackage{cite} %IEEE template package
\usepackage{amsmath,amssymb,amsfonts} %IEEE template package
\usepackage{algorithmic} %IEEE template package
\usepackage{graphicx} %IEEE template package
\usepackage{textcomp} %IEEE template package
\usepackage{xcolor} %IEEE template package
\usepackage{dblfloatfix} %enable figure placement other than the top of a page (default)

\definecolor{darkblue}{RGB}{0,40,86}
\usepackage[nolist]{acronym}
\usepackage{booktabs}
\usepackage{multirow}
\usepackage{etoolbox}

\def\BibTeX{{\rm B\kern-.05em{\sc i\kern-.025em b}\kern-.08em
    T\kern-.1667em\lower.7ex\hbox{E}\kern-.125emX}}

\begin{acronym}
  \acro{IoT}{Internet of Things}
\end{acronym}

\begin{document}

\title{
Edge-Deployable LLMs for Classification and Value-of-Information Ranking of Disaster-Related Tweets%
\thanks{Christopher Molek and K. Brent Venable were supported by NSF award NSF-RI-2008011.}
}

\author{
    \IEEEauthorblockN{
        Christopher D. Molek\IEEEauthorrefmark{1}\IEEEauthorrefmark{2},     
        K. Brent Venable\IEEEauthorrefmark{1}\IEEEauthorrefmark{2},
        Niranjan Suri\IEEEauthorrefmark{1}\IEEEauthorrefmark{2}\IEEEauthorrefmark{3}
        Stephen Russell\IEEEauthorrefmark{1}\\
    }
    
    \IEEEauthorblockA{
        \IEEEauthorrefmark{1} \textit{Department of Intelligent Systems \& Robotics - The University of West Florida (UWF), Pensacola, FL, USA}\\
        \textit{russell@uwf.edu}\\
        \IEEEauthorrefmark{2}\textit{Florida Institute for Human and Machine Cognition (IHMC), Pensacola, FL, USA}\\
        \textit{\{cmolek, bvenable, nsuri\}@ihmc.org}\\
        \IEEEauthorrefmark{3}\textit{US Army DEVCOM Army Research Laboratory (ARL), Adelphi, MD, USA} \\
        \textit{niranjan.suri.civ@mail.mil}\\
    }
}

\IEEEoverridecommandlockouts
\IEEEpubid{%
  \parbox{\columnwidth}{%
    \footnotesize
    \raggedright
    \copyright~2026 IEEE. Personal use of this material is permitted.
    Permission from IEEE must be obtained for all other uses,
    in any current or future media, including reprinting/republishing
    this material for advertising or promotional purposes, creating
    new collective works, for resale or redistribution to servers
    or lists, or reuse of any copyrighted component of this work
    in other works.
  }%
  \hspace{\columnsep}%
  \makebox[\columnwidth]{}%
}
\IEEEpubidadjcol %used to ensure no overlap with the copyright statement over the text on the first page
\maketitle

\begin{abstract}
Humanitarian-aid and disaster relief decisions often rely on noisy, incomplete, and uncertain information from social media. Although substantial research has focused on automating the classification of such data, limited work has examined whether individual posts contain sufficient information to support decisions to act. In this study, we use large language models (LLMs) to (1) classify social media content and (2) assess its Value-of-Information (VoI) in terms of operational actionability. We evaluate seven LLMs for the classification of 77,484 tweets across 10 humanitarian-aid categories and identify the Qwen3 (8B parameters) model as the top-performing classifier, achieving the highest F1 scores in all but one category. In a second task, the classified categories serve as contextual inputs for assessing VoI with respect to responder actionability across six well established metrics for each tweet. We observed that the Gemma 3 (4B parameters) model outperformed the other models on this task, achieving the highest recall across the majority of categories. This work demonstrates that relatively small, edge-deployable LLMs can prioritize actionable information in resource-constrained environments. By integrating VoI assessment, our framework advances the automated elicitation of high-utility knowledge from unstructured social media streams for disaster-response decision-making and highlights a pathway toward tactical edge applications using LLMs that fit on a single 8~GB GPU.
\end{abstract}

\begin{IEEEkeywords}
Value-of-Information, Actionability, Information sufficiency, Large Language Model, Decision Making, Disaster Response
\end{IEEEkeywords}

\section{Introduction}
Disaster response organizations must triage rapidly evolving, incomplete, and noisy social media streams to identify reports that inform operational decisions. Although classification organizes these streams, category labels alone fail to indicate whether a post contains decision-relevant detail to merit responder attention \cite{sunApplicationsArtificialIntelligence2020}.

Prior research established human-annotated benchmarks for humanitarian-aid classification and evaluated transformer and LLM performance on crisis-related social media tasks. Separately, actionability research primarily detects help requests and offers. However, it remains unclear whether edge-deployable LLMs can effectively zero-shot classify disaster-related posts using the same guidance as human annotation tasks 
%while concurrently 
as well as estimate their operational value.

This work addresses that gap by evaluating seven edge-deployable LLMs on the HumAID dataset for (1) humanitarian-aid category classification and (2) ETHANE-informed Value-of-Information (VoI) scoring. The ETHANE framework is an established approach used for emergency information elicitation and evaluates reports based on: \underline{E}xact Location, \underline{T}ype of Incident, \underline{H}azards, \underline{A}ccess to Scene, \underline{N}umber and Severity, and \underline{E}mergency Services \cite{jesipinteroperabilityboardJointDoctrineInteroperability2024}. 

Drawing on Zade \textit{et al.}'s finding that emergency responders view actionable information as ``anything which either they or their organization could use at that moment to assist, enact, or expedite the solution'' \cite{zadeSituationalAwarenessActionability2018}, we operationalize actionability-related VoI as the extent to which a post provides explicit, decision-relevant information aligned with the ETHANE framework.

Key contributions of this work include:
\begin{itemize}
\item Designing, implementing, and evaluating two fundamental components of  an LLM pipeline to automate categorization and actionability-based VoI assessment of disaster-related social media;

\item Creating an experimental framework to evaluate LLM alignment with human classifications and ETHANE-informed VoI scoring;
\item Assessing zero-shot, edge-deployable LLMs for classifying disaster-related tweets and computing ETHANE-informed VoI scores.
\end{itemize}

\section{Related Work}

The automation of information processing for humanitarian aid and disaster response has been an active area of research in computer science and artificial intelligence.
Alam \textit{et al.} collected roughly 24 million tweets and human-annotated 77,484 of them into 10 humanitarian-aid categories. This ``HumAID'' dataset spans 19 disaster events from 2016 to 2019 \cite{alamHumAIDHumanAnnotatedDisaster2021}. Their work established a benchmark for disaster-context annotation and classification, evaluating classical machine-learning, neural, and transformer-based approaches.

Recent work investigates LLM-based crisis classification via fine-tuning and zero-shot evaluation. Jinzhen \textit{et al.} demonstrated performance gains by fine-tuning Llama 3.1 8B-Instruct on HumAID, while McDaniel \textit{et al.} evaluated several LLMs in zero-shot crisis-triage settings, reporting performance variations based on model scale. These studies highlight LLM potential but leave open how smaller, edge-deployable models align with human annotation instructions on high-quality humanitarian-aid benchmarks \cite{jinzhenLightweightLLMFramework2026, mcdanielLLMsCrisisTriage2025, alamHumAIDHumanAnnotatedDisaster2021}. In contrast to these prior works, we formulate our prompts using the original HumAID Human Intelligence Task (HIT) instructions. Our approach, thus,  directly assesses how well LLM outputs align with human annotations.

Prior actionability research detects crisis-related content at varying granularities. Lamsal \textit{et al.} introduced the CrisisHelpOffer dataset and compact, encoder-only masked language models (reduced BERT architectures) to classify posts as requests, offers, both, or irrelevant \cite{lamsalActionableHelpCrises2025}. Their models identify explicit help-seeking and assistance-offering messages in resource-constrained environments.

Zguir \textit{et al.} extended request-and-offer detection using a fine-grained, multi-label taxonomy spanning supplies, actions, and emergency personnel \cite{zguirDetectingActionableRequests2025}. Their framework identifies specific resources, actions, and personnel while assigning a binary actionability label.

Our work addresses a complementary problem: assessing the extent to which a disaster-related tweet contains the information needed to support responder actionability. Rather than treating actionability merely as the presence of a request or offer, we operationalize this assessment using an ETHANE-informed Value-of-Information (VoI) score.

\section{Experimental design}
\label{e-des}

Our work focuses on  field deployment where cloud access may be intermittent or unavailable, and bandwidth may be reserved primarily for receiving incoming social-media data. For this reason 
we evaluate open-weight LLMs in the 1B--8B parameter range because they are compatible with resource-constrained edge systems.  Furthermore, we evaluate models without fine-tuning to establish a baseline based on pretrained knowledge, reflecting settings in which task-specific data, compute resources, time, or connectivity are insufficient for fine-tuning.

We selected models capable of operating on an 8~GB GPU and evaluated them using Ollama (v0.16.2) with Python (v3.12): Gemma2 (2B parameters), Llama3.2 (3B), Gemma3 (4B), Mistral-Instruct (7B), Llama3.1 (8B), DeepSeek-R1 (8B), and Qwen3 (8B). All models used 4-bit quantization; Gemma2 used Q4\_0, while the others used Q4\_K\_M. Experiments were conducted on a Dell XPS 15 with a 13th-generation Intel i9 processor (2.60~GHz), 32~GB RAM, and an Nvidia GeForce RTX 4060 Laptop GPU (8~GB). Most models were general-purpose open-weight LLMs selected to limit inference-time computational requirements, with DeepSeek-R1 and Qwen3 included as reasoning-enabled models for comparison.

% The majority of models were selected as simpler non-reasoning to keep compute resources low during inference. We did evaluate more complex models for comparison points to the simpler models.

\subsection{Experimental Design: Classification Task}
In this task, each tweet is presented to the LLMs alongside original human-labeling instructions \cite{alamCrisisMMDMultimodalTwitter2018, alamHumAIDHumanAnnotatedDisaster2021}. This enables direct evaluation of LLM–human alignment on a category-by-category basis, differing from the prior zero-shot crisis-triage evaluations \cite{gilardiChatGPTOutperformsCrowd2023}. The large language model (LLM) is tasked with assigning one category label to each tweet using a zero-shot prompting approach with role-based instructions. The prompt template for the classification task is presented below, Fig.~\ref{fig:classification_prompt_template}:

% In this task, each tweet is presented to the LLMs alongside a definition of humanitarian aid and a list of 10 predefined categories, each with a corresponding definition \cite{alamHumAIDHumanAnnotatedDisaster2021}. The large language model (LLM) is tasked with assigning one category label to each tweet using a zero-shot prompting approach with role-based instructions. The prompt template for the classification task is presented below, Fig.~\ref{fig:classification_prompt_template}:

% For the classification task, we replicate the original human-labeling instructions \cite{alamCrisisMMDMultimodalTwitter2018, alamHumAIDHumanAnnotatedDisaster2021}. Providing LLMs with the exact Human Intelligence Task (HIT) descriptions enables a direct evaluation of LLM–human alignment \cite{gilardiChatGPTOutperformsCrowd2023}. Unlike prior zero-shot crisis-triage evaluations, we assess alignment with the HumAID annotation instructions on a category-by-category basis and extend categorization to a second, operationally focused task.
\begin{figure}[ht]
\centering
\footnotesize
\setlength{\fboxsep}{5pt}
\setlength{\fboxrule}{0.5pt}

\fbox{%
\begin{minipage}{0.95\columnwidth}
{\centering\textbf{Humanitarian-Aid Classification Prompt Template}\par}

\vspace{1mm}
\hrule
\vspace{1mm}

\textbf{Humanitarian-aid definition}\\
\textit{[Humanitarian-aid definition inserted here]}

\vspace{1mm}

\textbf{Humanitarian-aid categories}\\
\textit{[Category names and corresponding definitions inserted here]}

\vspace{1mm}
\hrule
\vspace{1mm}

\textbf{Tweet}\\
\texttt{RT @globeandmail: Red Cross to transfer \$50-million to evacuees from Fort McMurray wildfire}

\vspace{1mm}
\hrule
\vspace{1mm}

\textbf{Instruction}\\
You are an information categorizer and need to pick one sort category to apply to the tweet.

\vspace{1mm}

\textbf{Required JSON response format}

\vspace{1mm}

{\ttfamily
\begin{tabular}{@{}l@{}}
\{\\
\quad "your\_logic": "your reasoning here",\\
\quad "category": "your picked category"\\
\}
\end{tabular}
}

\end{minipage}%
}
\caption{Prompt template for zero-shot humanitarian-aid
classification. The humanitarian-aid definition and category definitions remain fixed across tweets; only the tweet text changes for each inference.}
\label{fig:classification_prompt_template}
\end{figure}
The categories that  we consider in this work are consistent with those identified in \cite{alamCrisisMMDMultimodalTwitter2018} and correspond to: 
"Caution and Advice",
"Displaced People and Evacuations",
"Infrastructure and Utility Damage",
"Injured or Dead People",
 "Missing or Found People",  
"Not Humanitarian", 
"Other Relevant Information",
"Requests for Urgent Needs",
"Rescue, Volunteering or Donation Effort", and, "Sympathy and Support".  
In evaluating each tweet only the tweet text changes. All models in the classification task receive the same prompt for a given tweet and use a temperature setting of zero.

\subsection{Experimental Design: VoI Scoring Task}\label{sec:ED_VOI}
Following classification, the humanitarian-aid category provides decision-maker context for ETHANE-informed questions that extract actionability-relevant information and produce a VoI score. We restrict the VoI analysis to three categories: ``Requests or Urgent Needs'', ``Injured or Dead People'', and ``Infrastructure and Utility Damage'', selected for semantic distinctiveness, definition succinctness, and representation of different classification-performance levels (see Fig.~\ref{fig:f1_heat_map}).
% Following classification, we extend the analysis to incorporate the concept of VoI. The category establishes the decision-maker context for ETHANE-informed questions, which extract actionability-relevant information to compute the VoI score. 

% We restricted the VoI analysis to three categories: ``Requests or Urgent Needs'', ``Injured or Dead People'', and ``Infrastructure and Utility Damage''. These were chosen based on their semantic distinctiveness, the succinctness of their definitions as well as being representative of different performance classes for the classification task (see Fig. \ref{fig:f1_heat_map}). 

For each category, we construct a six-question binary rubric informed by the United Kingdom Joint Decision Making Model and ETHANE reporting framework \cite{jesipinteroperabilityboardJointDoctrineInteroperability2024}, with responses scored as ``Stated'' or ``Not Stated''. The resulting ETHANE-informed VoI score is the number of rubric elements explicitly stated in a tweet, ranging from 0 to 6. Category definitions provide contextual guidance consistent with Zade \textit{et al.}'s concept of actionability.
% Each rubric consists of six binary questions scored as either ``Stated'' or ``Not Stated''. The resulting ETHANE-informed VoI score represents the number of rubric elements explicitly stated in a tweet, ranging from 0 to 6.

% To assign value, we use category definitions as contextual guidance alongside Zade \textit{et al.}'s definition of actionability: whether immediate action can be taken based on the information provided in a tweet. To elicit this information, we follow the United Kingdom's implementation of the Joint Decision Making Model and  we develop a rubric consisting of six binary evaluation questions informed by the ETHANE reporting framework \cite{jesipinteroperabilityboardJointDoctrineInteroperability2024}.

In our formulation, the humanitarian-aid category provides responder context corresponding most closely to ``\underline{E}mergency Services Reporting'', so no separate question is used for this ETHANE component. An additional \underline{N}umber and Severity question improves rubric fidelity by capturing category-specific information relevant to ETHANE reporting. Figure~\ref{fig:ethane_rubric} shows the rubric for ``Requests or Urgent Needs''.
% In our formulation, the humanitarian-aid category provides responder context and most closely relates to the final ETHANE component, ``\underline{E}mergency Services Reporting.'' We therefore use the category label as contextual information rather than formulating a separate question for this component. An additional \underline{N}umber and Severity question improves rubric fidelity and captures category-specific information relevant to this component of ETHANE reporting.
% An example of the ETHANE-informed rubric for the ``Requests or Urgent Needs'' category is shown in Fig.~\ref{fig:ethane_rubric}.
\begin{figure}[ht]
\centering
\footnotesize
\setlength{\fboxsep}{5pt}
\setlength{\fboxrule}{0.5pt}

\fbox{%
\begin{minipage}{0.95\columnwidth}
\textit{``Requests or Urgent Needs'' – Reports of urgent needs or supplies such as food, water, clothing, money, medical supplies, blood, or evacuation.}
\vspace{1mm}
\hrule
\vspace{1mm}

\begin{enumerate}
    \item Is the location where aid is needed mentioned (e.g., neighborhood, street, building, city)?
    \item Does the tweet say what kind of incident caused the need (e.g., flooding, earthquake, fire, landslide)?
    \item Are any stated safety hazards mentioned preventing aid workers delivering supplies (e.g. unstable debris, gas leak)?
    \item Does the tweet describe how aid workers can access or leave the area (e.g., roads blocked, bridge out, only boat access)?
    \item Does the tweet state what is needed (e.g., water, food, shelter, evacuation, medical supplies)?
    \item Does the tweet state how many people need help or supplies (even approximately)?
\end{enumerate}

\end{minipage}%
}
\caption{ETHANE-informed rubric for ``Requests or Urgent Needs'' category.}
\label{fig:ethane_rubric}
\end{figure}

The ETHANE framework was used to generate similar questions for two other  categories based on their definitions: “Injured or Dead People” – Reports of injured, trapped, or dead people due to the disaster; “Infrastructure and Utility Damage” – Reports of damage to infrastructure such as buildings, houses, roads, bridges, power lines, communication poles, or vehicles.

In the rubric each question is associated with a score of 1 when the relevant information is explicitly stated in the tweet and 0 otherwise. Higher total scores indicate greater utility for decision-making, particularly with respect to enabling operational action. A tweet containing a complete report with respect to its category receives a score of six.

The prompt template for the VoI scoring task is presented in Fig.~\ref{fig:voi_prompt_template}. This prompt changes with the tweet text, within each aid category. The category definition and corresponding rubric questions change based on the category associated to the tweet. All models receive the same prompt for a given tweet, except for the specified tweet and category changes, and use a temperature setting of zero.

\begin{figure}[!b]
\centering
\footnotesize
\setlength{\fboxsep}{5pt}
\setlength{\fboxrule}{0.5pt}

\fbox{%
\begin{minipage}{0.95\columnwidth}
{\centering\textbf{ETHANE-Informed VoI Prompt Template}\par}

\vspace{1mm}
\hrule
\vspace{1mm}

\textbf{Instructions}

\begin{enumerate}
    \item Read the tweet and the humanitarian-aid category definition.
    \item Answer each rubric question using \textit{only} information explicitly stated in the tweet.
    \item If the information is not mentioned in the tweet, select \textit{Not stated}.
    \item Do not guess or infer from general disaster context, hashtags, or unstated assumptions.
    \item Tweets may be incomplete sentences or truncated; answer the questions to the best of your ability.
\end{enumerate}

\vspace{1mm}
\hrule
\vspace{1mm}

\textbf{Category context}\\
\textit{[Humanitarian-aid category and category definition inserted here]}

\vspace{1mm}

\textbf{Tweet}\\
\texttt{Please bring awareness and help @LOHANTHONY}

\vspace{1mm}

\textbf{Category-specific ETHANE-informed rubric}\\
\textit{[Six category-specific questions inserted here]}

\vspace{1mm}
\hrule
\vspace{1mm}

\textbf{Required JSON response format}

\vspace{1mm}

{\ttfamily
\begin{tabular}{@{}l@{}}
\{\\
\quad "question\_1\_logic": "your reasoning here",\\
\quad "question\_1\_answer": "Stated or Not Stated",\\
\quad \ldots\\
\quad "question\_6\_logic": "your reasoning here",\\
\quad "question\_6\_answer": "Stated or Not Stated"\\
\}
\end{tabular}
}

\end{minipage}%
}
\caption{Prompt template for the ETHANE-informed VoI scoring. The humanitarian-aid category definition and six rubric questions vary by aid category; the tweet text varies for every inference.}
\label{fig:voi_prompt_template}
\end{figure}

Reference VoI scores were generated using GPT-4o by applying the same category-specific ETHANE-informed rubric to each tweet. We validated this approach against human annotations for 60 tweets (20 per category). The primary author independently annotated the subset before comparing those annotations with the GPT-4o outputs. Across 360 binary rubric judgments, GPT-4o agreed with 356, yielding 98.9\% question-level agreement. Based on this agreement, GPT-4o outputs were used as reference scores for the full dataset, against which the smaller LLMs were evaluated.

Within this pipeline, decision-makers can define a threshold parameter \(t = n\) to filter out tweets obtaining a score below a given VoI threshold score, thereby mitigating information overload. Additionally, to prevent error propagation from classification mistakes, this study uses tweets with human-labeled HumAID categories.
%as context for the ETHANE-informed VoI scoring task.

\section{Results and Discussion}
In this section we illustrate and discuss the experimental results for both tasks. 
\subsection{Classification Task - Results and Discussion}
\begin{figure}[htbp]
\centerline{\includegraphics[width=1\columnwidth]{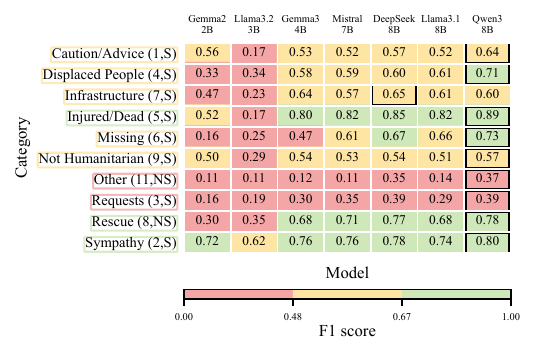}}
\caption{Classification F1 scores for each model. Results fall into three performance bands indicating strong (green), moderate (yellow), and weak (red) alignment with human labels.}
\label{fig:f1_heat_map}
\end{figure}

We first compare model outputs for the classification tasks against the human-labeled ground truth from Alam \textit{et al.} \cite{alamHumAIDHumanAnnotatedDisaster2021}. 
Fig.~\ref{fig:f1_heat_map} shows the F1 scores related to this comparison where  black boxes indicate top scores. Qwen3 outperforms the others in all but one category, ``Infrastructure and Utility Damage''. It achieves an F1 score of 0.894 for ``Injured or Dead People'', demonstrating strong precision, recall, and alignment with human labels. Conversely, Qwen3 struggles with ``Requests or Urgent Needs'', scoring just 0.394 and reflecting weak alignment with human annotations.

Category-specific median F1 scores reveal three broad performance bands regarding human alignment: strong (\(\geq\)0.67), moderate (0.48--0.66), and weak (\(<\)0.48). Green outlines of the category names in Fig. \ref{fig:f1_heat_map} denote categories where the median is strongly aligned: ``Injured or Dead People'', ``Rescue, Volunteering, or Donation Effort'', and ``Sympathy and Support''. 
Similarly, yellow and red  outlines denote moderate and weak  alignment,  respectively.
%categories: ``Caution and Advice'', ``Displaced People and Evacuations'', ``Infrastructure and Utility Damage'', ``Missing or Found People'', and ``Not Humanitarian''. Red outlines indicate weakly aligned categories: ``Other Relevant Information'' and ``Requests or Urgent Needs''.

To explain low performance on these categories, we examine two factors: (1) the prompt presentation order of category and definition, and (2) the succinctness of each category definition.

We find no clear relationship between the position of the category in the prompt (denoted by the number next to the category name in Fig.~\ref{fig:f1_heat_map}) and classification performance. For instance, two categories for which there was strong alignment-- ``Sympathy and Support'' and "Rescue, Volunteering, or Donation Effort'' -- appeared, near the beginning and end, respectively.
Yet, both are associated with similar scores. Similarly, definition succinctness fails to explain performance patterns. We marked definitions as succinct (S) or not succinct (NS) in Fig.~\ref{fig:f1_heat_map}. The broad, ambiguous ``Other Relevant Information'' category and the highly specific ``Requests or Urgent Needs'' category both yield similarly low F1 scores. Therefore, neither order nor succinctness of category definition explain the performance gaps.

Analyzing precision and recall provides deeper insight. For ``Requests or Urgent Needs'', recall consistently exceeds precision. Gemma2, for example, achieves 0.819 recall but only 0.090 precision. This pattern indicates models correctly identify relevant instances but generate numerous false positives, dragging down precision. Conversely, for ``Other Relevant Information'', recall trails precision, suggesting models struggle to identify relevant instances. Qwen3, for instance, achieves 0.436 precision and 0.328 recall. Although this analysis leaves the root cause unclear, it demonstrates the models fail on these categories in fundamentally different ways.

\subsection{VoI Scoring Task - Results and Discussion}

To evaluate the performance of our target LLMs in the VoI scoring task, we compare the models' outputs against the human-validated GPT-4o reference scores obtained as described in Section~\ref{e-des}. We measure alignment as agreement on individual binary questions. We, then, define tweet-level agreement as  the percentage of tweets aligning with the reference scores on at least \(k\) of six questions (\(k \in \{1,\ldots,6\}\)). These values are cumulative, therefore alignment percentages increase as \(k\) decreases.

\begin{figure}[htbp]
    \centering
    \includegraphics[
        width=\columnwidth
    ]{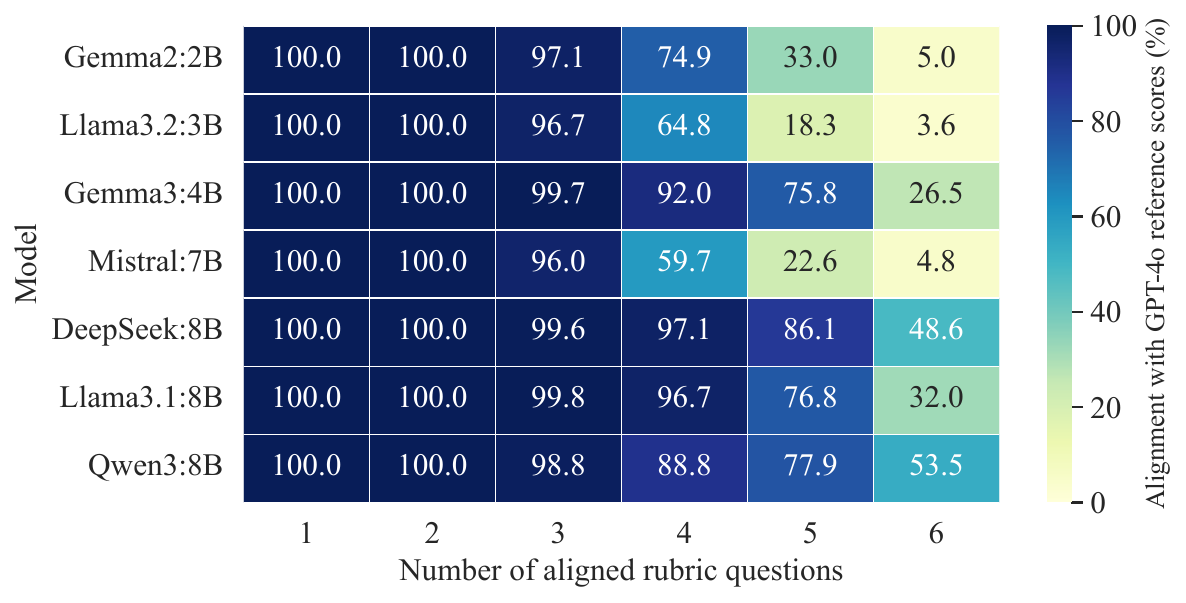}
    \caption{Percentage of tweets for ``Injured or Dead People'' aligning with GPT-4o reference scores at each number of aligned rubric questions.}
    \label{fig:heat_map_alignment_IDP}
\end{figure}

Across all three categories, most models achieve high alignment with GPT-4o reference scores when relaxing the criterion to four aligned rubric questions. The ``Injured or Dead People'' category exhibits the lowest four-question alignment and serves to illustrate model-level differences (Fig.~\ref{fig:heat_map_alignment_IDP}).

For the ``Injured or Dead People'' category, four models---Gemma3, DeepSeek, Llama3.1, and Qwen3---achieve greater than 75\% alignment on at least five of six questions, spanning a 10.3 percentage point range (Fig.~\ref{fig:heat_map_alignment_IDP}). DeepSeek achieves the highest five-question alignment at 86.1\%, whereas Gemma3, Llama3.1, and Qwen3 perform similarly. These same four models also demonstrate strong alignment at the four-question threshold, scoring at least 88.8\% (spanning an 8.3 percentage point range). 

For the ``Requests or Urgent Needs'' category, we observed  Mistral, DeepSeek, Llama3.1, and Qwen3 each aligning with the reference scores on at least five of six questions for over 82\% of tweets, differing by no more than 5.3 percentage points. At the four-question threshold, all models except Gemma2 (85.4\%) exceed 94\% alignment, with Gemma3, Mistral, DeepSeek, Llama3.1, and Qwen3 spanning a 3.1 percentage point range.

For the ``Infrastructure and Utility Damage'' category, DeepSeek and Qwen3 exceed 70\% alignment at the five-question threshold, whereas all other models achieve at least 56.3\%. At the four-question threshold, all models except Mistral (87.1\%) achieve at least 88\% alignment. Overall, most models align on at least four of six rubric questions for a large majority of tweets across all categories. However, alignment drops substantially when tightening the criterion from four to five questions, demonstrating that complete or near-complete agreement across all rubric elements remains challenging.

Across the three categories, Qwen3 achieves the highest complete-alignment scores (all six rubric questions): 50.8\% in ``Requests or Urgent Needs'' and 42.4\% in ``Infrastructure and Utility Damage''. Furthermore, most models more than double their complete-alignment performance when relaxing the criterion from six to five aligned questions. Qwen3 (across the three categories) and DeepSeek (for ``Injured or Dead People'') are exceptions because both already achieve relatively high complete-alignment baseline scores. 
%%%%%%%%%% 
%Note placement of this figure needs to be before the first text of the page you want it on to ensure the figure ends up at the bottom of the page and not the bottom of the following page.  Also the preamble must have:
% \usepackage{dblfloatfix} %enable figure placement other than the top of a page (default)
%%%%%%%%%%
\begin{figure*}[h]
    \centering
    \includegraphics[
        width=\textwidth
    ]{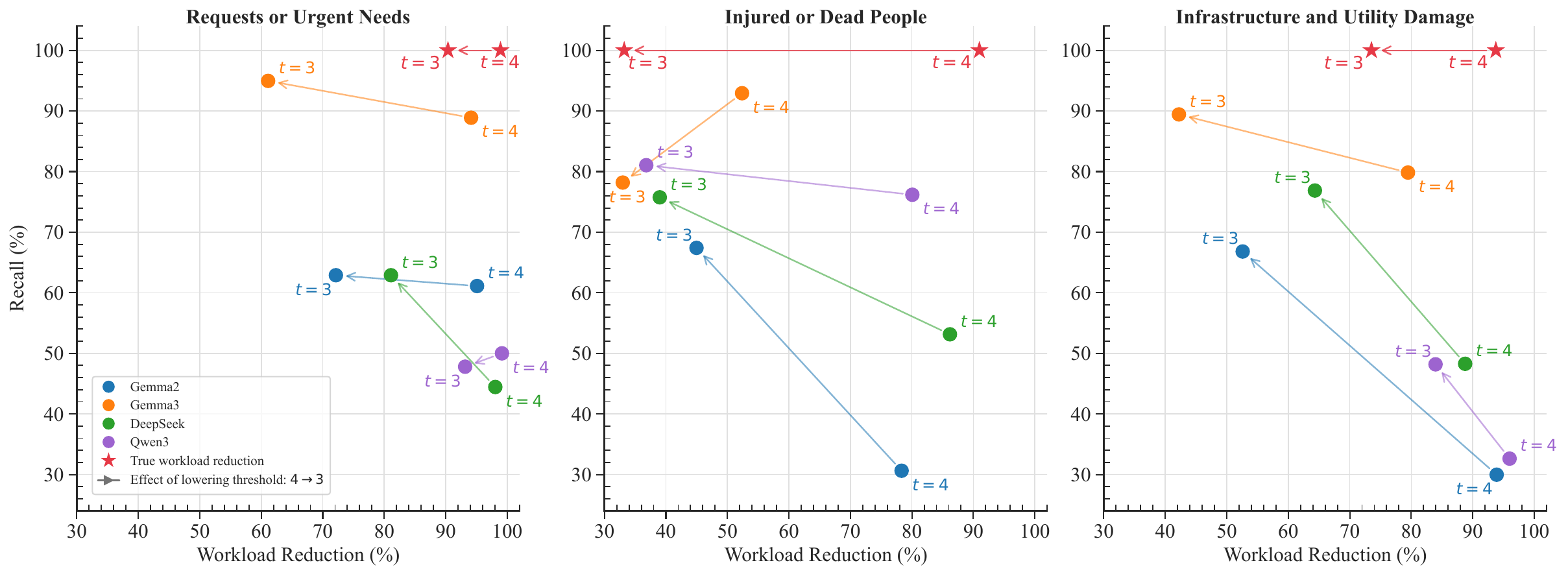}
    \caption{Recall versus workload reduction. Two VoI threshold conditions are shown: \(t=3\) and \(t=4\).}
    \label{fig:recall_work_reduced_mcc}
\end{figure*}

In the proposed approach, decision-makers specify a minimum VoI threshold, \(t\), forwarding only tweets with a predicted score of at least \(t\). Table~\ref{tab:voi_gpt4o_distributions} summarizes the distribution of GPT-4o reference VoI scores across categories, with values in parenthesis indicating total tweet counts. We evaluate operating thresholds of \(t=4\) and \(t=3\), requiring at least four and three explicit rubric elements, respectively. The \(t=4\) threshold serves as our baseline because alignment performance drops rapidly beyond this point.

\begin{table}[htbp]
\caption{Distribution of GPT-4o reference VoI scores per category.}
\begin{center}
\begin{tabular}{cccc}
\toprule
VoI Score & \thead{Requests or \\Urgent Needs\\(1651)} & \thead{Injured or \\Dead People\\(4236)} & \thead{Infrastructure and \\Utility Damage\\(6045)} \\
\midrule
0 & 410 & 160 & 899 \\
1 & 637 & 288 & 1529 \\
2 & 445 & 959 & 2017 \\
3 & 141 & 2447 & 1223 \\
4 & 18 & 379 & 337 \\
5 & 0 & 3 & 39 \\
6 & 0 & 0 & 1 \\
\bottomrule
\end{tabular}
\label{tab:voi_gpt4o_distributions}
\end{center}
\end{table}

% \begin{figure*}[htpb]
%     \centering
%     \includegraphics[
%         width=\textwidth
%     ]{voi_selected_models_workload_reduction_recall_reference_benchmarks.pdf}
%     \caption{Recall versus workload reduction. Two VoI threshold conditions are shown: \(t=3\) and \(t=4\).}
%     \label{fig:recall_work_reduced_mcc}
% \end{figure*}

To evaluate model performance, we measure recall, workload reduction, and the Matthews correlation coefficient (MCC), an established measure of quality for binary classification in the presence of imbalanced classes. In disaster response, missing high-VoI tweets carries severe operational costs, thus maximizing recall is our primary objective. Workload reduction percentage is calculated as \((1-((TP+FP)/N))*100\), where \(N\) represents total tweet volume and higher values indicate greater filtering selectivity. Fig.~\ref{fig:recall_work_reduced_mcc} illustrates recall versus workload reduction across select models. For visual clarity, we depict only the top-performing models in recall and MCC -- Gemma2, Gemma3, DeepSeek, and Qwen3. In Fig.~\ref{fig:recall_work_reduced_mcc}, star symbols represent the true workload reduction based on the reference score distributions in Table~\ref{tab:voi_gpt4o_distributions}. Effective models achieve high recall alongside a workload reduction near this benchmark, positioning them toward the upper-right region of each plot. Finally, to account for class imbalance across categories and threshold values, we report and analyze MCC scores alongside these metrics.

For ``Requests or Urgent Needs'' at \(t=4\), Gemma3 recovered 16 of 18 reference-positive tweets (88.9\% recall) while forwarding 97 of 1,651 tweets, reducing workload by 94.1\%. While other models achieved similar workload reductions, they suffered significantly lower recall (Fig.~\ref{fig:recall_work_reduced_mcc}, left). Qwen3 achieved the highest MCC score (0.56) and Gemma3 the second highest (0.37). At \(t=3\), instead, Gemma3 identified 151 of 159 reference-positive tweets (eight false negatives) but forwarded 642 tweets (38.9\% of the category), quadrupling the workload from the true workload. While other models reduced workload more effectively, all produced over three times as many false negatives as Gemma3. MCC scores for all models remained within 0.04 of their respective \(t=4\) values, indicating similar predictive capabilities between threshold levels when the class balance changes.

For ``Injured or Dead People'' at \(t=4\), Gemma3 has the lowest FN count (27), recovering 355 of 382 reference-positive tweets (92.9\% recall) while forwarding 2,017 tweets (47.6\% of the category; Fig.~\ref{fig:recall_work_reduced_mcc}, center). Qwen3 forwarded far fewer tweets (844; 19.9\% of the category) but missed 91 reference-positive tweets (76.2\% recall), illustrating a distinct workload--recall trade-off between these two models. Qwen3 has the highest MCC score (0.44), whereas Gemma3 has a score of 0.29.
For this category,  at \(t=3\), Qwen3 achieved the lowest false-negative count among models, identifying 2,293 of 2,829 reference-positive tweets (81.1\% recall) while forwarding 2,677 tweets. Notably, the true workload reduction drops sharply from 91.0\% at \(t=4\) to 33.2\% at \(t=3\), revealing a significant number of tweets containing exactly three rubric elements. This unique distribution, at \(t=3\), suggests a need for category refinement—for example, separating injured and deceased individuals, whose operational urgency and response teams differ. MCC scores for three of the four models, excluding Gemma2, remained within 0.08 of their respective \(t=4\) values, indicating similar predictive capabilities between threshold levels when the class balance changes. Gemma2 changes 0.28, which is undesirable to have a performance change this large for a changing threshold.

For ``Infrastructure and Utility Damage'' at \(t=4\), Gemma3 achieved the highest recall (79.8\%), recovering 301 of 377 reference-positive tweets while forwarding 1,241 tweets (79.5\% workload reduction). DeepSeek and Qwen3 were more selective: DeepSeek forwarded 679 tweets (48.3\% recall; 195 false negatives), while Qwen3 forwarded 242 tweets (32.6\% recall; 254 false negatives). Gemma3 and Qwen3 score the highest MCC score at 0.38. At \(t=3\), Gemma3 reached 89.4\% recall (1,431 of 1,600 reference-positive tweets) but forwarded 3,493 tweets (57.8\% of the category). DeepSeek offered superior workload reduction at this threshold, forwarding 2,156 tweets (35.7\% of the category) while maintaining 76.9\% recall. Here both Qwen3 and DeepSeek exhibit shifts in MCC scores of 0.15 and 0.21, respectively. Gemma2 and Gemma3 both stay within 0.02 with changing threshold.

\section{Concluding Remarks, Limitations, and Future Work}

\textit{Concluding Remarks:}
For humanitarian-aid tweet classification, Qwen3 achieved the strongest overall performance, while Gemma3 performed comparably to Mistral, DeepSeek, and Llama3.1. Llama3.2 produced the lowest overall classification performance. Our ETHANE-informed VoI framework significantly reduces reviewer workload while identifying information needed to support actionability. Model alignment with reference VoI scores declines substantially when more than four rubric questions must be aligned. Among compact models, Gemma3 demonstrated the strongest performance, rivaling 8B models, while DeepSeek maintained competitive VoI alignment despite comparatively weak classification performance.

\textit{Limitations:}
Although this work establishes an end-to-end pipeline, we intentionally evaluate its components independently to isolate errors at each stage rather than evaluating the fully integrated pipeline. In addition, the current framework equally weights the six information elements. Practical implementations will likely require unequal weighting because some elements may be essential for action; for example, responders may be unable to act without location information. 
The resulting VoI score reflects the extent to which a tweet contains the information needed to support actionability; however, the present evaluation does not establish whether a given score corresponds to a tweet being operationally actionable.  
The framework also does not yet assess whether these elements exceed an information-sufficiency threshold or integrate information across tweets to determine whether accumulated information is sufficient to change a responder's decision. Establishing this would require follow-up studies involving realistic operational scenarios and human responders.

Lastly, although the models are evaluated on a resource-constrained, edge-capable platform with an 8~GB GPU, we do not characterize performance in terms of latency, throughput, or power consumption. These measurements are reserved for future work.

\textit{Future Work:}
Future work will evaluate the integrated pipeline with predicted classifications feeding the VoI assessment and investigate classification improvements through insights gained from lexical and semantic analyses of tweet text. To address the human-annotation limitation, we will collect human-annotated datasets using the ETHANE-informed rubrics to evaluate alignment between LLM-generated VoI assessments and human assessments. 

The framework will also be extended to consider unequal information weighting, source credibility, information freshness, and integration across multiple tweets to form a more complete, timely, and credible operational picture. Further study is needed to determine which information elements, and how many, are necessary to influence a decision maker's course of action.

% Because missing critical disaster reports can have serious operational consequences, recall may need to be prioritized over workload reduction. Across all three categories, Gemma3 was the most robust to threshold-induced class imbalance, providing superior recall in nearly every instance at the cost of higher review workload. Conversely, Qwen3 achieved the greatest workload reduction but lower recall in two of the three categories. These results demonstrate that a compact 4B model such as Gemma3 can provide effective edge-based VoI screening when minimizing missed high-utility information is the primary objective.
% For VoI assessment, minimizing false negatives is particularly important for high-consequence categories such as Injured or Dead People'', where missing a severe injury report may pose greater operational risk than missing a shelter request in ``Requests or Urgent Needs''. Future work will collect human-annotated datasets using the ETHANE-informed rubrics to evaluate alignment between LLM-generated VoI assessments and human assessments.

% \section*{Acknowledgment}

\bibliographystyle{IEEEtran}
\bibliography{references}

\end{document}